\documentclass{elsart}
\usepackage{natbib}
\usepackage{epsfig}
\begin{document}
\runauthor{Alejandro Gangui, Levon Pogosian, Serge Winitzki}
\begin{frontmatter}

\title{Cosmic String Signatures in the Cosmic Microwave Background
Anisotropies\thanksref{Sometwo}}
\author[Paestum,Baiae]{Alejandro Gangui\thanksref{Someone}}
\author[Imperial]{Levon Pogosian}
\author[Tufts]{Serge Winitzki}
\address[Paestum]{Instituto de Astronom\'{\i}a y F\'{\i}sica del Espacio,
         Ciudad Universitaria, 1428 Buenos Aires, Argentina}
\address[Baiae]{Departamento de F\'{\i}sica, Universidad de Buenos Aires,
         Ciudad Universitaria -- Pab. 1, 1428 Buenos Aires, Argentina}
\address[Imperial]{Theoretical Physics, The Blackett Laboratory,
         Imperial College, Prince Consort Road, London SW7 2BZ, United Kingdom}
\address[Tufts]{Department of Physics and Astronomy,
                Tufts University, Medford, MA 02155, USA}
\thanks[Sometwo]{New Astronomy Reviews, in press (2002).}
\thanks[Someone]{Member of CONICET, partially supported by Fundaci\'on Antorchas.}

\newcommand{\be}{\begin{equation}}
\newcommand{\ee}{\end{equation}}
\newcommand{\dy}{\dot{y}}
\newcommand{\sun}{{\rm Sun}}

\begin{abstract}
We briefly review certain aspects of cosmic microwave background
anisotropies as generated in passive and active models of structure
formation.  We then focus on cosmic strings based models and discuss
their status in the light of current high-resolution observations from
the BOOMERanG, MAXIMA and DASI collaborations.  Upcoming megapixel
experiments will have the potential to look for non-Gaussian features
in the CMB temperature maps with unprecedented accuracy. We therefore
devote the last part of this review to treat the non-Gaussianity of
the microwave background and present a method for computation of the
bispectrum from simulated string realizations.
\end{abstract}

\end{frontmatter}

\section{Introduction}
\typeout{SET RUN AUTHOR to \@runauthor}

Anisotropies of the Cosmic Microwave Background radiation (CMB) are
directly related to the origin of structure in the universe. Galaxies
and clusters of galaxies eventually formed by gravitational
instability from primordial density fluctuations, and these same
fluctuations left their imprint on the CMB. Recent balloon
\cite{boomerang,maxima} and ground-based interferometer \cite{dasi}
experiments have produced reliable estimates of the power spectrum of
the CMB temperature anisotropies. While they helped eliminate certain
candidate theories for the primary source of cosmic perturbations, the
power spectrum data are still compatible with theoretical
estimates of a relatively large variety of models, such as
$\Lambda$CDM, quintessence models or some hybrid models including
cosmic defects.  These models, however, differ in their predictions
for the statistical distribution of the anisotropies beyond the power
spectrum.  The MAP (currently in space) and Planck (scheduled for launch in 2007)
satellite missions will provide high-precision data allowing definite
estimates of non-Gaussian signals in the CMB. It is therefore
important to know precisely what are the predictions of all candidate
models for the statistical quantities that will be extracted from the
new data, and to identify specific signatures of the various models.

There are two main classes of models of structure
formation---\textit{passive} and \textit{active} models. In passive
models, density inhomogeneities are set as initial conditions at some
early time, and while they subsequently evolve as described by
Einstein-Boltzmann equations, no additional perturbations are
seeded. On the other hand, in active models a time-dependent source
generates new density perturbations through all time.

Most realizations of passive models are based on the idea of
inflation. In simplest inflationary models it is assumed that there
exists a weakly coupled scalar field $\phi$, called the inflaton,
which ``drives'' the (approximately) exponential expansion of the
universe. Quantum fluctuations of $\phi$ are stretched by the
expansion to scales beyond the horizon, thus ``freezing'' their
amplitude.  Inflation is followed by a period of thermalization,
or reheating,
during which energy is transferred to usual particles. Because
of the spatial variations of $\phi$ introduced by quantum
fluctuations, thermalization occurs at slightly different times in
different parts of the universe. Such fluctuations in the
thermalization time give rise to density fluctuations. Because of
their quantum nature and because of the fact that initial
perturbations are assumed to be in the vacuum state and hence well
described by a Gaussian distribution, perturbations produced during
inflation are expected to follow Gaussian statistics to a high degree,
or either be products of Gaussian random variables. This is a fairly
general prediction that will be tested shortly with MAP and more
thoroughly in the future with Planck data.

Active models of structure formation are motivated by cosmic
topological defects with the most promising candidates being cosmic
strings. It is widely believed that the universe underwent a series of
phase transitions as it cooled down due to the expansion. If our ideas
about grand unification are correct, then some cosmic defects, such as
domain walls, strings, monopoles or textures, should have formed
during phase transitions in the early universe \cite{vilshel}. Once
formed, cosmic strings could survive long enough to seed density
perturbations.  Generically, perturbations produced by active models
are not expected to be Gaussian.

Defect models possess the attractive feature that they have no
parameter freedom, as all the necessary information is in principle
contained in the underlying particle physics model. This is a feature
that fascinated Dennis Sciama: during his days at Trieste, he would
always be willing to discuss these ``marvelous cosmic defects'' with
everyone. He even encouraged a few students {\em sans liaisons
contraignantes} to pursue studies on this promising subject and
facilitated all necessary means for it.  The relevant, ubiquitous
dimensionless quantity for every astrophysical and cosmological
signature left by strings is $G\mu$, where $G$ is Newton's constant
and $\mu$ stands for the mass per unit length of the string, with a
value of roughly $G\mu \sim 10^{-6}$ for GUT strings. This happens to
be just the correct order of magnitude of the level of CMB
anisotropies. Back in 1996, the power spectrum plot of $l(l+1)C_l$
vs.~$l$ consisted of some scattered points and strings were considered
a likely candidate for the origin of structure formation.  Moreover,
the error bars were so big that Dennis would never miss the
opportunity to tease inflation fans who wanted to see a peak in that
plot: ``Come on'', he would say, ``you do not claim you actually {\em
see\/} a peak there, do you..!''.

To close this section, let us mention that, in addition to purely
active or passive scenarios, perturbations could be seeded by some
combination of the two mechanisms. For example, cosmic strings could
have formed just before the end of inflation and partially contributed
to seeding density fluctuations. It has been shown
\cite{hybrid} that such hybrid models are very successful in
fitting the CMB power spectrum data. Cosmic strings are generally
expected to produce distinguishing non-Gaussian features in the CMB
and it will soon become possible to look for them in the data from MAP
and Planck.

\section{CMB power spectrum from strings}

Calculating CMB anisotropies sourced by topological defects is a
rather difficult task. In inflationary scenario the entire information
about the seeds is contained in the initial conditions for the
perturbations in the metric.  In the case of cosmic defects,
perturbations are continuously seeded over the period of time from the
phase transition that had produced them until today. The exact
determination of the resulting anisotropy requires, in principle, the
knowledge of the energy-momentum tensor [or, if only two point
functions are being calculated, the unequal time correlators
\cite{uetc}] of the defect network and the products of its decay at
all times. This information is simply not available! Instead, a number
of clever simplifications, based on the expected properties of the
defect networks ({\it e.g.} scaling), are used to calculate the
source. The latest data from BOOMERanG~\cite{boomerang},
MAXIMA~\cite{maxima} and DASI~\cite{dasi} experiments clearly disagree
with the predictions of these simple models of defects~\cite{dgs96}.

The shape of the CMB angular power spectrum is determined by three
main factors: the geometry of the universe, coherence and causality.

The curvature of the universe directly affects the paths of light rays
coming to us from the surface of last scattering \cite{weinberg}. In a
closed universe, because of the lensing effect induced by the positive
curvature, the same physical distances between points on the sky would
correspond to larger angular scales. As a result, the peak structure
in the CMB angular power spectrum would shift to the larger angular
scales or the smaller values of $l$.

\begin{figure}[tbp]
\centerline{\epsfxsize = 0.6
\hsize
\epsfbox{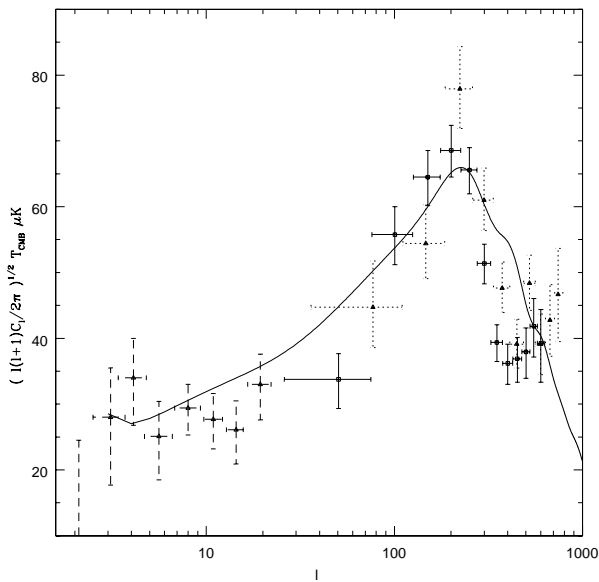}}
\vskip 0.1 truecm
\caption{Data from  BOOMERanG (solid error bars), MAXIMA (dotted error bars) and
DASI (dashed error bars). The line shows the CMB power spectrum produced
by the wiggly local string model of Ref. \protect \cite{PV99} in a closed universe with
$\Omega_{\rm total}=1.3$, $\Omega_{\rm baryon}=0.05$, $\Omega_{\rm CDM}=0.35$,
$\Omega_{\Lambda}=0.9$, and $H_0 = 65 {\rm ~km ~s}^{-1} {\rm Mpc}^{-1}$.}
\label{omega}
\end{figure}

The prediction of the cosmic string model of \cite{PV99} for
$\Omega_{\rm total}=1.3$ is shown in Fig.~\ref{omega}.
As can be seen, the main peak in the angular power spectrum can be
matched by choosing a reasonable value for $\Omega_{\rm
total}$. However, even with the main peak in the right place the
agreement with the data is far from satisfactory. The peak is
significantly wider than that in the data and there is no sign of a
rise in power at $l\approx 600-800$ suggested by the data. The
situation is even worse for global strings which fail to produce a
distinct peak of any kind.  For local strings, the sharpness and the
height of the main peak in the angular spectrum can be enhanced by
including the effects of the gravitational radiation \cite{hind98} and
wiggles \cite{PV99}. More precise high-resolution numerical
simulations of string networks in realistic cosmologies with a large
contribution from $\Omega_{\Lambda}$ are needed to determine the exact
amount of small-scale structure on the strings and the nature of the
products of their decay. It is, however, unlikely that including these
effects alone would result in a sufficiently narrow main peak or any
presence of secondary peaks\footnote{The statistical significance of
the observed secondary peaks in the CMB angular spectrum is still an
open issue, see {e.g.} Ref.~\cite{pedro01}.}.
It seems that now Dennis would not have had strong arguments to tease his friends.

Let us next consider issues of causality and coherence and how the
random nature of the string networks comes into the calculation of the
anisotropy spectrum.

Both experimental and theoretical results for the CMB power spectra
involve calculations of averages. To estimate the correlations of
the observed temperature anisotropies, experimentalists compute the
average over all available patches on the sky. When calculating the
predictions of their models, theorists find the average over the
{\em ensemble} of possible outcomes provided by the model.

In inflationary models, as in all passive models, only the initial
conditions for the perturbations are random.  The subsequent evolution
is the same for all members of the ensemble. For wavelengths higher
than the Hubble radius, the linear evolution equations for the Fourier
components of such perturbations have a growing and a decaying
solution. The modes corresponding to smaller wavelengths have only
oscillating solutions. As a consequence, prior to entering the
horizon, each mode undergoes a period of phase ``squeezing'' which
leaves it in a highly coherent state by the time it starts to
oscillate. Coherence here means that all members of the ensemble,
corresponding to the same Fourier mode, have the same temporal
phase. So even though there is randomness involved, as one has to draw
random amplitudes for the oscillations of a given mode, the time
behavior of different members of the ensemble is highly
correlated. The total power spectrum is an ensemble averaged
superposition of all Fourier modes and the coherence results in an
interference pattern seen as the acoustic peaks.

\begin{figure}[tbp]
\centerline{\epsfxsize = 0.7
\hsize
\epsfbox{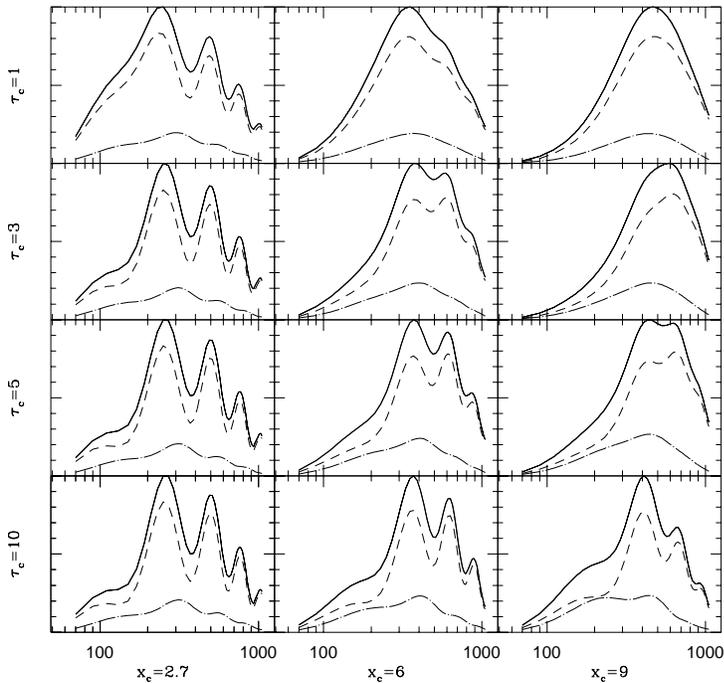}} \vskip 0.5 truecm
\caption{Predictions of the toy model of Magueijo {\it et al.}
\protect \cite{mag&alb} for different values of parameters $x_c$, the
coherence length, and $\tau_c$, the coherence time; $x_c\propto \eta /
\lambda_c(\eta)$, where $\eta$ is the conformal time and
$\lambda_c(\eta)$ is the correlation length of the network at time
$\eta$. One can obtain oscillations in the CMB power spectrum by
fixing either one of the parameters and varying the other. }
\label{magetal}
\end{figure}

In contrast, the evolution of the string network is highly
non-linear. Cosmic strings are expected to move at relativistic
speeds, self-intersect and reconnect in a chaotic fashion. A
consequence of this behavior is that the unequal time correlators of
the string energy-momentum vanish for time differences larger than a
certain coherence time. Members of the ensemble corresponding to a
given mode of perturbations will have random temporal phases with the
``dice'' thrown on average once in each coherence time. The coherence
time of a realistic string network is rather short. As a result, the
interference pattern in the angular power spectrum is completely
washed out.

Causality manifests itself, first of all, through the initial
conditions for the string sources, the perturbations in the metric and
the densities of different particle species. If one assumes that
defects are formed by a causal mechanism in an otherwise smooth
universe, then the correct initial conditions are obtained by setting
the components of the stress-energy pseudo-tensor $\tau_{\mu \nu}$ to
zero \cite{veersteb,PenSperTur}. These are the same as the
isocurvature initial conditions \cite{HuSperWhite}. A generic
prediction of isocurvature models (assuming perfect coherence) is that
the first acoustic peak is almost completely hidden. The main peak is
then the second acoustic peak and in flat geometries it appears at
$l\approx 300-400$. This is due to the fact that after entering the
horizon a given Fourier mode of the source perturbation requires time
to induce perturbations in the photon density.

Causality also implies that no superhorizon correlations in the string
energy density are allowed. The correlation length of a ``realistic''
string network is normally between 0.1 and 0.4 of the horizon size.

An interesting study was performed by Magueijo {\it et al.}
\cite{mag&alb} where they have constructed a toy model of defects with
two parameters: the coherence length and the coherence time. The
coherence length was taken to be the scale at which the energy density
power spectrum of the strings turns from a power law decay for large
values of $k$ into a white noise at low $k$. This is essentially the
scale corresponding to the correlation length of the string
network. The coherence time was defined in the sense described in the
beginning of this section, in particular, as the time difference
needed for the unequal time-correlators to vanish.  Their study showed
(see Fig.~\ref{magetal}) that by accepting any value for one of the
parameters and varying the other (within the constraints imposed by
causality) one could reproduce the oscillations in the CMB power
spectrum. Unfortunately for cosmic strings, at least as we know them
today, they fall into the parameter range corresponding to the upper
right corner in Fig.~\ref{magetal}.  In order to fit the
observations, cosmic strings must either be more coherent or they have
to be stretched over larger distances, which is another way of making
them more coherent. To understand this, imagine that there were only one
long straight string stretching across the universe and moving with some
velocity. The evolution of this string would be linear and the induced
perturbations in the photon density would be coherent. By increasing
the correlation length of the string network we would move closer to
this limiting case of just one long straight string and so the
coherence would be enhanced.

\begin{figure}[tbp]
\centerline{\epsfxsize = 0.7
\hsize
\epsfbox{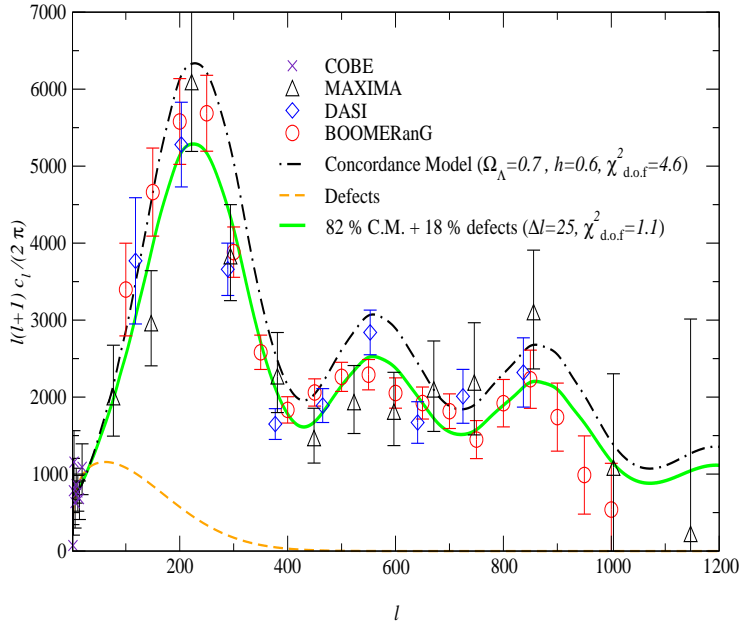}}
\vskip 0.1 truecm
\caption{$\ell (\ell + 1) C_\ell$ versus $\ell$ for three different
models [the figure is borrowed from Bouchet {et al.}, Ref. \protect
\cite{hybrid}].  The upper (dot-dashed) line represents the prediction
of a $\Lambda$CDM model, with $n_{\rm S} = 1$, $\Omega_\Lambda = 0.7$,
$\Omega_{\rm m} = 0.3$, $\Omega_{\rm b} = 0.05$ and $h = 0.6$.  The
lower dashed line is a typical spectrum produced by global strings.
Both lines are normalized at the COBE scale (crosses).  Combining both
curves, with their relative contribution as a free parameter, produces
the solid curve. The string contribution is about 18\% of the total.}
\label{mixed}
\end{figure}

The question of whether or not defects can produce a pattern of
the CMB power spectrum similar to, and including the peaks of, that
produced by the adiabatic inflationary models was repeatedly addressed
in the literature \cite{hybrid,mag&alb,liddle,turok,varyingc}. In
particular, it was shown \cite{mag&alb,turok} that one can construct a
causal model of active seeds which for certain values of parameters
can reproduce the oscillations in the CMB spectrum. The main problem
today is that current realistic models of cosmic strings fall out of
the parameter range that is needed to fit the observations.

In addition to purely active or passive models, perturbations could be
seeded by some combination of the two mechanisms \cite{hybrid}.
Figure~\ref{mixed} (borrowed from Bouchet {et al.}) shows two
uncorrelated spectra produced by an inflationary $\Lambda$CDM model
(dot-dashed line) and a global string model (dashed line), both
normalized on the COBE data, as well as a weighted sum of the two
(solid line).  The best fit to the latest BOOMERanG, MAXIMA and DASI
data 
is obtained with a non-negligible string contribution of $18$\%. At
the moment, models combining strings with inflation \cite{hybrid} or
models of varying speed of light \cite{varyingc} are the only ones
involving topological defects that to some extent can match the
observations. One possible way to distinguish their predictions from
those of inflationary models would be to compute a non-Gaussian
statistic, such as the CMB bispectrum.

\section{CMB Bispectrum from strings}

\subsection{The string model}

Let us now turn to a computation of the bispectrum of CMB
temperature anisotropies as generated by simulated cosmic strings.
To calculate the sources of perturbations we use an updated version of
the wiggly string model of Ref.\cite{PV99} (see also \cite{ABR97}).

The network is represented by a collection of uncorrelated, straight
string segments moving with random, uncorrelated velocities, and are
produced at some early epoch. At every subsequent epoch, a certain
fraction of the number of segments decays in a way that maintains
network scaling.  The length of each segment at any time is taken to
be equal to the correlation length of the network which, together with
the root mean square velocity of segments, are computed from the
velocity-dependent one-scale model of Martins and Shellard
\cite{MS00,MS96}. The positions of segments are drawn from a uniform
distribution in space and their orientations are chosen from a uniform
distribution on a two-sphere.

According to the one-scale model \cite{Kib,Ben,MS96,MS00}, the string
network at any time can be characterized by a single length scale, the
correlation length $L$, defined by
$
\rho={\mu}/{L^2}$,
where $\rho$ is the energy density in the string network.
Expansion of space stretches the strings without altering their mass per
unit length. If simply stretched in that manner, strings would
soon come to dominate the energy density of the universe. However,
at the same time with the expansion, long strings reconnect and
chop off loops which later decay. These two competing effects
result in an intermediate steady-state (the so-called ``scaling
solution'') in which the characteristic scale of the string
network remains constant relative to the horizon size.

The total energy of the string network in a volume $V$ at any time
$\tau$ is
\begin{equation}
N \mu L = V\rho=\frac{\mu V}{L^2} \label{vrho}
\end{equation}
where $N=N(\tau )$ is the total number of string segments at that time, $V=V_0 a^3$, $a$ is the expansion factor
and $V_0$ is a constant simulation volume. From Eq.(\ref{vrho}) it follows that
$N={V}/{L^3}={V_0}/{{\ell}^3}$,
where we have defined ${\ell} \equiv L/a$. The comoving length
${\ell}$ is approximately proportional to the conformal time $\tau$
and implies that the number of strings $N(\tau)$ within the simulation
volume $V_0$ falls as $\tau^{-3}$. To calculate the CMB anisotropy we
need to evolve the string network over at least four orders of
magnitude in cosmic expansion. Hence we would have to start with $N
\geq 10^{12}$ string segments in order to have one segment left at the
present time. Keeping track of such a large number of segments is
numerically impossible. Fortunately, a way around this difficulty was
suggested in Ref. \cite{ABR97}.

The suggestion is to consolidate all string segments that decay at the
same epoch. The number of segments that decay at the (discretized) conformal time
$\tau_i$ is
\begin{equation}
N_d (\tau_i ) = V [ n(\tau_{i-1}) - n(\tau_i) ] \label{eq:nd}
\end{equation}
where $n(\tau )=[\ell(\tau )]^{-3}$ is the number density of strings
at time $\tau $. Then, with the formalism we employ, instead of
summing over $\sum _{i=1}^{K}N_{d}(\tau_{i})\geq
10^{12}$ segments we now sum over only $K$ segments, making $K$ a
parameter. Such a simplification only works for computation of two- and three-point
correlation functions under the assumption that string segments are statistically
independent. We refer the interested reader to our recent article~\cite{usthree},
where we explain the procedure in more detail.

As a final comment, let us note that the simulation model in this form
does not allow computation of CMB sky maps. This is because the method
of finding the two- and three-point functions we describe involves
{}``consolidated{}'' quantities which do not correspond to the
energy-momentum tensor of a real string network. These quantities are
auxiliary and specially prepared to give the correct two- or
three-point functions after ensemble averaging.  Having clarified
this, the stress-energy from the simulated string model can be found
and used to compute the appropriate density perturbations which lead
to the expected CMB temperature anisotropies.

\subsection{Computation of bispectrum from string realizations}

It is conventional to expand CMB temperature fluctuations into spherical harmonics:
\begin{equation}
{\Delta T / T} ({\bf
e})=\sum _{l m} a_{l m} Y_{l m}({\bf e}).
\end{equation}
The coefficients $a_{lm}$ can be decomposed into Fourier modes,
\begin{equation}
\label{eq:alm-def}
a_{lm} = \frac{V_0}{(2\pi)^3} \left( -i\right)^{l}4\pi
\int d^{3}{\mathbf k}\, \Delta_{l}\left( {\mathbf k}\right)
Y^{*}_{lm}(\hat{{\mathbf k}}) .
\end{equation}
Given the sources $\Theta_{\mu\nu}({\mathbf k},\tau)$
produced by the string model, $\Delta _{l}( {\mathbf k})$ are found by
solving linearized Einstein-Boltzmann equations and integrating along
the line of sight \cite{cmbfast}.  This procedure can be written as
the action of a linear operator ${\hat B}_l^{\mu\nu}(k)$ on the source
energy-momentum tensor, $ \Delta _{l}({\mathbf k}) ={\hat
B}_l^{\mu\nu}(k) \Theta_{\mu\nu}({\mathbf k},\tau)$.
The third moment of
$a_{lm}$'s
can be expressed as
\begin{eqnarray}
\nonumber
\left\langle
a_{l_{1}m_{1}}a_{l_{2}m_{2}}a_{l_{3}m_{3}}\right\rangle &=& \left(
-i\right) ^{l_{1}+l_{2}+l_{3}}\left( 4\pi \right)^{3}
\!\!\frac{V_0^3}{(2\pi)^9}\!\!\int\!\! d^{3}{\mathbf
k}_{1}d^{3}{\mathbf k}_{2}d^{3}{\mathbf k}_{3} Y^{*}_{l_{1}m_{1}}\!(
\hat{{\mathbf k}}_{1}) \\
&\times& Y^{*}_{l_{2}m_{2}}\!( \hat{{\mathbf k}}_{2}) Y^{*}_{l_{3}m_{3}}\!( \hat{{\mathbf k}}_{3}) \left\langle
\Delta_{l_{1}}\!\!\left( {\mathbf k}_{1}\right) \Delta_{l_{2}}\!\!\left( {\mathbf k}_{2}\right)
\Delta_{l_{3}}\!\!\left( {\mathbf k}_{3}\right) \right\rangle . \label{eq:3alm-1}
\end{eqnarray}
Because of homogeneity we have
\begin{eqnarray}
& & \left\langle \Delta _{l_{1}}\left( {\mathbf
k}_{1}\right) \Delta _{l_{2}}\left( {\mathbf k}_{2}\right) \Delta
_{l_{3}}\left( {\mathbf k}_{3}\right) \right\rangle =
\delta^{(3)} \left( {\mathbf k}_{1}+{\mathbf k}_{2}+{\mathbf
k}_{3}\right) P_{l_1 l_2 l_3}\left( {\mathbf k}_{1},{\mathbf
k}_{2},{\mathbf k}_{3}\right).
\label{p3lvector}
\end{eqnarray}
Given scalar values $k_{1}$, $k_{2}$, $k_{3}$, there is a unique (up
to an overall rotation) triplet of directions $\hat{{\mathbf k}}_{i}$
for which the RHS of Eq.~(\ref{p3lvector}) does not vanish. The
quantity $P_{l_1 l_2 l_3}\left( {\mathbf k}_{1},{\mathbf
k}_{2},{\mathbf k}_{3}\right)$ is invariant under an overall rotation
of all three vectors ${\mathbf k}_{i}$ and therefore may be
equivalently represented by a function of scalar values $k_{1}$,
$k_{2}$, $k_{3}$. Hence, we get
\begin{eqnarray}
& & \left\langle \Delta _{l_{1}}\left( {\mathbf
k}_{1}\right) \Delta _{l_{2}}\left( {\mathbf k}_{2}\right) \Delta
_{l_{3}}\left( {\mathbf k}_{3}\right) \right\rangle =
\delta^{(3)} \left( {\mathbf k}_{1}+{\mathbf k}_{2}+{\mathbf
k}_{3}\right) P_{l_1 l_2 l_3} (k_1,k_{2},k_{3}).
\label{p3lscalar}
\end{eqnarray}
The triangle constraint ${\mathbf k}_{1}+{\mathbf k}_{2}+{\mathbf
k}_{3} = 0$ must be satisfied. Then, in terms of the simulation volume
$V_0$, we have
\begin{equation}
P_{l_1 l_2 l_3} \!\left( k_{1},k_{2},k_{3}\right) \! =\!
\frac{(2\pi)^3}{V_0}\left\langle \Delta_{l_{1}}\!\!\left( {\mathbf
k}_{1}\right) \Delta_{l_{2}}\!\!\left( {\mathbf k}_{2}\right)
\Delta_{l_{3}}\!\!\left( {\mathbf k}_{3}\right) \right\rangle
. \label{p3l0}
\end{equation}
Given an arbitrary direction $\hat{\mathbf{k}}_1$ and the magnitudes
$k_1$, $k_2$ and $k_3$, the directions $\hat{\mathbf{k}}_2$ and
$\hat{\mathbf{k}}_3$ are specified up to rotations by the triangle
constraint.  Both sides of Eq. (\ref{p3l0}) are function of scalar
$k_i$ only. The expression on the RHS will be computed numerically by
averaging over different realizations of the string network {\it and}
over all directions $\hat{\mathbf{k}}_1$. Because of isotropy and
since the allowed sets of directions $\hat{\mathbf{k}}_i$ are planar,
it is enough to restrict the numerical calculation to directions
$\hat{\mathbf{k}}_i$ in a fixed plane.  This significantly reduces the
computational expense.

Substituting Eqs. (\ref{p3lvector}) to (\ref{p3l0}) into
(\ref{eq:3alm-1}), Fourier transforming the Dirac delta and using the
Rayleigh identity, we can perform all angular integrations
analytically and obtain a compact form for the third moment
\begin{equation}
\label{eq:3alm-res}
\left\langle a_{l_{1}m_{1}}a_{l_{2}m_{2}}a_{l_{3}m_{3}}\right\rangle
= {\mathcal H}_{l_{1}l_{2}l_{3}}^{m_{1}m_{2}m_{3}} \int r^{2}dr \, b_{l_1 l_2 l_3} (r)  ,
\end{equation}
where, noting $\left(^{\,\, l_1~\,\;l_2~\,\;l_3}_{m_1~m_2~m_3}\right)$ a Wigner 3$j$-symbol, we have
\begin{eqnarray}
\label{eq:hlll}
{\mathcal H}_{l_{1}l_{2}l_{3}}^{m_{1}m_{2}m_{3}} &\equiv & \sqrt{\frac{\left(
2l_{1}+1\right) \left( 2l_{2}+1\right) \left( 2l_{3}+1\right) }{4\pi
}} \left(
\begin{array}{ccc}
l_{1} & l_{2} & l_{3}\\
0 & 0 & 0
\end{array}\right)
\left( \begin{array}{ccc}
l_{1} & l_{2} & l_{3}\\
m_{1} & m_{2} & m_{3}
\end{array}\right) \, ,
\end{eqnarray}
and where we have defined, in terms of spherical Bessel functions $j_{l}$,
\begin{eqnarray}
\nonumber b_{l_1 l_2 l_3} (r) 
&\equiv& 
\frac{8}{\pi^3} \frac{V_0^3}{(2\pi)^3} 
\int k_1^{2}dk_1\, k_2^{2}dk_2 \, k_3^{2}dk_3 \,
\\ &\times& 
j_{l_1} (k_1 r) j_{l_2} (k_2 r) j_{l_3} (k_3 r) P_{l_1 l_2 l_3} (k_1,k_{2},k_{3}) .
\label{defineb}
\end{eqnarray}

Our numerical procedure consists of computing the RHS of
Eq.~(\ref{eq:3alm-res}) by evaluating the necessary integrals and
averaging over many realizations of the strings sources.
Readers wishing more details can consult Ref.~\cite{usthree}.

From Eq.~(\ref{eq:3alm-res}) we easily derive the CMB
bispectrum ${\mathcal C}_{l_1 l_2 l_3 }$, defined as \cite{GM00}
\begin{eqnarray}
\bigl\langle a_{l_1 m_1} a_{l_2 m_2} a_{l_3 m_3} \bigr\rangle =
\left( \begin{array}{ccc}
l_{1} & l_{2} & l_{3}\\
m_{1} & m_{2} & m_{3}
\end{array}\right)
{\mathcal C}_{l_1 l_2 l_3 } \, .
\end{eqnarray}
We now restrict our calculations to the angular bispectrum
$C_{l_{1}l_{2}l_{3}}$ in the `diagonal' case,
\textit{i.e.}~$l_{1}=l_{2}=l_{3}=l$.  This is a representative case
and, in fact, the one most frequently considered in the
literature. The power spectrum is usually plotted in terms of
$l(l+1)C_{l}$ which, apart from constant factors, is the contribution
to the mean squared anisotropy of temperature fluctuations per unit
logarithmic interval of $l$. In full analogy with this, the relevant
quantity to work with in the case of the bispectrum is
\begin{eqnarray} G_{lll} = l(2l+1)^{3/2}\left( \begin{array}{ccc}
l & l & l\\
0 & 0 & 0
\end{array}\right) C_{lll}\, .\label{eq:QtP}
\end{eqnarray}
For large values of the multipole index $l$, $G_{lll}\propto
l^{3/2}C_{lll}$.  Note also what happens with the 3$ j$-symbols
appearing in the definition of the coefficients
${\mathcal{H}}_{l_{1}l_{2}l_{3}}^{m_{1}m_{2}m_{3}}$: the symbol
$\left( ^{\, \, l_{1}\, \; l_{2}\, \; l_{3}}_{m_{1}m_{2}m_{3}}\right)
$ is absent from the definition of $C_{l_{1}l_{2}l_{3}}$, while in
Eq.~(\ref{eq:QtP}) the symbol $\left( ^{\, l\; l\; l}_{0\: 0\:
0}\right) $ is squared. Hence, there are no remnant oscillations due
to the alternating sign of $\left( ^{\, l\; l\; l}_{0\: 0\: 0}\right)$.

However, even more important than the value of $C_{lll}$ itself is the
relation between the bispectrum and the cosmic variance associated
with it.  In fact, it is their comparison that tells us about the
observability `in principle' of the non-Gaussian signal. The cosmic
variance constitutes a theoretical uncertainty for all observable
quantities and is due to having just one
realization of the stochastic process, in our case, the CMB sky.

The way to proceed is to employ an estimator
$\hat{C}_{l_{1}l_{2}l_{3}}$ for the bispectrum and compute the
variance from it. By choosing an unbiased estimator we ensure it
satisfies $C_{l_{1}l_{2}l_{3}}=\langle
\hat{C}_{l_{1}l_{2}l_{3}}\rangle $. However, this condition does not
isolate a unique estimator. The proper way to select the {\it best
unbiased} estimator is to compute the variances of all candidates and
choose the one with the smallest value.  The estimator with this
property was computed in Ref.~\cite{GM00} and is \begin{equation}
\label{eq:clll-best} \hat{C}_{l_{1}l_{2}l_{3}}=\! \! \! \sum
_{m_{1},m_{2},m_{3}}\! \!  \left(
\begin{array}{ccc} l_{1} & l_{2} & l_{3}\\
m_{1} & m_{2} & m_{3}
\end{array}\right) a_{l_{1}m_{1}}a_{l_{2}m_{2}}a_{l_{3}m_{3}}.
\end{equation}
The variance of this estimator, assuming a mildly non-Gaussian
distribution, can be expressed in terms of the angular power spectrum
$C_{l}$ as follows~\cite{GM00}
\begin{equation}
\label{eq:sigma}
 \sigma
^{2}_{\hat{C}_{l_{1}l_{2}l_{3}}}\! \! \! \!
=C_{l_{1}}C_{l_{2}}C_{l_{3}}\!
\left( 1\! +\! \delta _{l_{1}l_{2}}\! \! +\! \delta _{l_{2}l_{3}}\! \!
+\!
\delta _{l_{3}l_{1}}\! \! +\! 2\delta _{l_{1}l_{2}}\delta
_{l_{2}l_{3}}\right) .
\end{equation}
The theoretical signal-to-noise ratio for the bispectrum (in the
diagonal case $l_{1}=l_{2}=l_{3}=l$) is then given by
\begin{equation}
(S/N)_{l_{1}l_{2}l_{3}} =
|C_{l_{1}l_{2}l_{3}}/\sigma_{\hat{C}_{l_{1}l_{2}l_{3}}}|.
\end{equation}
Incorporating all the specifics of the particular experiment, such as
sky coverage, angular resolution, etc., will allow us to give an
estimate of the particular non-Gaussian signature associated with a
given active source and, if observable, indicate the appropriate range
of multipole $l$'s where it is best to look for it.

\begin{figure}
{\par\centering \leavevmode\epsfig{file=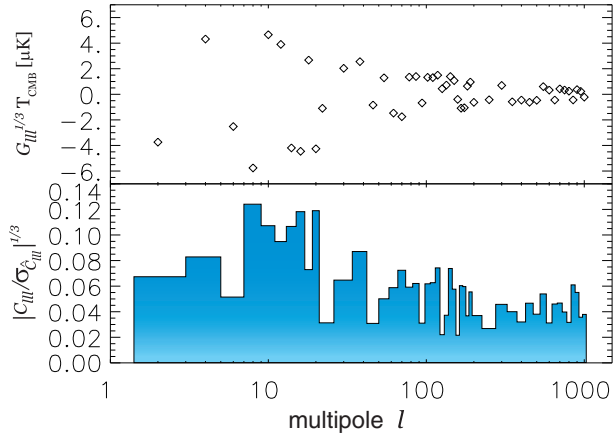, width=9cm}
\par}
\caption{The CMB angular bispectrum in the `diagonal' case
($G_{lll}^{1/3}$) from wiggly cosmic strings in a spatially flat model
with cosmological parameters $\Omega _{\textrm{CDM}}=0.3$,
$ \Omega_{\textrm{baryon}}=0.05$, $ \Omega _{\Lambda }=0.65$, and Hubble
constant $ H=0.65{\textrm{km}}{\textrm{s}}^{-1}{\textrm{Mpc}}^{-1}$
{[}upper panel{]}. In the lower panel we show the ratio of the signal
to theoretical noise $ |C_{lll}/\sigma _{\hat{C}_{lll}}|^{1/3}$ for
different multipole indices.
Normalization follows from fitting the power spectrum to the
BOOMERanG and MAXIMA data.}
\label{fig:1}
\end{figure}

In Fig.~\ref{fig:1} we show the results for $G_{lll}^{1/3}$
{[}cf.~Eq.~(\ref{eq:QtP}){]}. It was calculated using the string model
with $800$ consolidated segments in a flat universe with cold dark
matter and a cosmological constant. Only the scalar contribution to
the anisotropy has been included. Vector and tensor contributions are
known to be relatively insignificant for local cosmic strings and can
safely be ignored in this model \cite{ABR97,PV99}\footnote{The
contribution of vector and tensor modes is large in the case of global
strings \cite{uetc,dgs96}.}.  The plots are produced using a single
realization of the string network by averaging over $720$ directions
of ${\mathbf k}_{i}$. The comparison of $G_{lll}^{1/3}$ (or
equivalently ${C}_{lll}^{1/3}$) with its cosmic variance
{[}cf.~Eq.~(\ref{eq:sigma}){]} clearly shows that the bispectrum (as
computed from our cosmic string model) lies hidden in the theoretical
noise and is therefore undetectable for any given value of $l$.

Let us note, however, that in its present stage our string code
describes Brownian, wiggly long strings in spite of the fact that long
strings are very likely not Brownian on the smallest scales.  In
addition, the presence of small string loops \cite{proty} and
gravitational radiation into which they decay were not yet included in
our model. These are important effects that could, in principle,
change our predictions for the string-generated CMB bispectrum on very
small angular scales.

\section{Outlook}

The imprint of cosmic strings on the CMB is a combination of different
effects.  Prior to the time of recombination strings induce density
and velocity fluctuations on the surrounding matter. During the period
of last scattering these fluctuations are imprinted on the CMB through
the Sachs-Wolfe effect: namely, temperature fluctuations arise because
relic photons encounter a gravitational potential with spatially
dependent depth. In addition to the Sachs-Wolfe effect, moving long
strings drag the surrounding plasma and produce velocity fields that
cause temperature anisotropies due to Doppler shifts.  While a string
segment by itself is a highly non-Gaussian object, fluctuations
induced by string segments before recombination are a superposition of
effects of many random strings stirring the primordial plasma. These
fluctuations are thus expected to be Gaussian as a result of the
central limit theorem.

As the universe becomes transparent, strings continue to leave their
imprint on the CMB mainly due to the Kaiser-Stebbins effect.  This
effect creates line discontinuities in the temperature field of
photons passing on opposite sides of a moving long string.  However,
this effect can produce non-Gaussian perturbations only on
sufficiently small scales. This is because on scales larger than the
characteristic inter-string separation at the time of the
radiation-matter equality, the CMB temperature perturbations result
from superposition of effects of many strings and are likely to be
Gaussian.

Currently, strongest constraints on cosmic string models come from the
CMB power spectrum data. Predictions of strings-only models are
inconsistent with data from BOOMERanG, MAXIMA and DASI. However, the
data can still be fitted reasonably well by ``hybrid'' models that
combine strings and adiabatic inflationary perturbations as sources
for primordial fluctuations. Presence of cosmic strings could, in
principle, be distinguished by non-Gaussian statistics.  The new
high-resolution CMB measurements from MAP and PLANCK will allow a
precise estimate of non-Gaussian signals in the CMB.

The most straightforward non-Gaussian discriminator is the three-point
function of temperature, or equivalently (in Fourier space) the
angular bispectrum. In Ref.~\cite{usthree}, we have developed a
numerical method for computing the CMB angular bispectrum from first
principles in any active model of structure formation, such as cosmic
defects, where the energy-momentum tensor is known or can be
simulated. The method does not use CMB sky maps and requires a
moderate amount of computations. We applied this method to the
computation of some relevant components of the bispectrum produced
from a model of cosmic strings and did not find a non-Gaussian signal
that would be observable in the forthcoming satellite-based CMB
missions. This indicates that either the model we used to simulate
cosmic strings is inadequate for extracting the non-Gaussian
signature, or that the angular bispectrum is not an appropriate
statistic for detecting cosmic strings.

Avelino \textit{et al.}~\cite{ASWA98} applied several non-Gaussian
tests to the perturbations seeded by cosmic strings. They found the
density field distribution to be close to Gaussian on scales larger
than $1.5 (\Omega _M h^2)^{-1}$ Mpc, where $\Omega _M$ is the fraction
of cosmological matter density in baryons and CDM combined. Scales
this small correspond to the multipole index of order $l \sim
10^4$. We have not attempted a calculation the CMB bispectrum on these
scales because the linear approximation is almost guaranteed to fail
at such small scales, and because of increased computational cost for
higher $l$ multipoles.

The interest in cosmic strings has considerably declined since they
have been ruled out as the primary seed for structure formation in the
universe. This, however, is not at all sufficient to rule out strings
as an interesting topic of research. Studying formation, evolution,
properties and cosmological implications of cosmics strings and other
topological defects remains to be both exciting and important. Certain
cosmological puzzles, such as baryogenesis and ultra-high energy
cosmic rays, could be resolved with the help of topological
defects~\cite{vilshel}. Recent experimental and theoretical progress
in understanding properties of topological defects in low-temperature
condensed matter systems, such as superfluid helium and
superconductors, has led to an intriguing possibility of studying
Cosmology in the Lab \cite{grisha}. Any new experimental constraint on
the population and properties of cosmic defects is potentially
relevant to building viable cosmological and particle physics
models. Since current and future CMB experiments will provide the most
precise cosmological data, it is absolutely essential to use them
maximally in learning as much as we can about cosmic defects.

\vspace{2cm}

\hrule

\vspace{2cm}

\begin{figure}[htbp]
  \begin{center}
    \leavevmode
    \epsfxsize = 5cm
    \epsffile{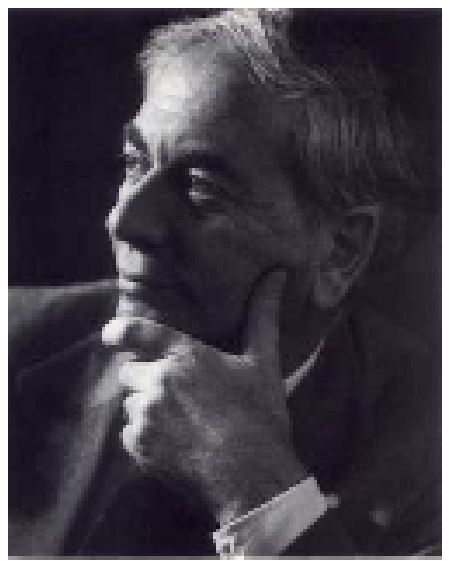}
  \end{center}
\end{figure}           

\begin{center}
{\em   me voy por un camino de labores fecundas, \\
       a mi muerte yunques ta\~ned, callad campanas 
       \footnote{I'm leaving down a road of fruitful labours, \\ 
           $~~~$ anvils toll at my death, hush, bells} } 
\end{center}


\vspace{2cm}

\hrule

\end{document}